\documentclass[preprint,floatfix,eqsecnum,aps,nofootinbib]{revtex4}

\usepackage{amsmath}
\usepackage{amssymb}
\usepackage{amscd}
\usepackage{mathrsfs}
\usepackage{graphicx}
\usepackage{subfigure}
\usepackage{slashed}
\usepackage{mathbbol}
\usepackage{subfigure}
\usepackage{float}

\begin{document}

\title{BPS solitons in Lifshitz field theories}
\author{Archil Kobakhidze}\email{archilk@unimelb.edu.au}
\affiliation{School of Physics, The University of Melbourne, Victoria 3010, Australia}
\author{Jayne E. Thompson}\email{j.thompson@pgrad.unimelb.edu.au}
\affiliation{School of Physics, The University of Melbourne, Victoria 3010, Australia}
\author{Raymond R. Volkas}\email{raymondv@unimelb.edu.au}
\affiliation{School of Physics, The University of Melbourne, Victoria 3010, Australia}

\begin{abstract}
Lorentz-invariant scalar field theories in $d+1$ dimensions
with second-order derivative terms are unable to support static soliton
solutions that are both finite in energy and stable for $d>2$, a result known as Derrick's theorem.
Lifshitz theories, which introduce
higher-order spatial derivatives, need not obey Derrick's theorem.  We construct
stable, finite-energy, static soliton solutions in Lifshitz scalar field theories
in $3+1$ dimensions with dynamical critical exponent $z=2$.  We exhibit three generic types: non-topological
point defects, topological point defects, and topological strings.
We focus mainly on Lifshitz theories that are defined through a superpotential and admit
BPS solutions.  These kinds of theories are the bosonic sectors of supersymmetric theories derived from the
stochastic dynamics of a scalar field theory in one higher dimension.
If nature obeys a Lifshitz field theory in the ultraviolet, then the novel topological defects discussed here may
exist as relics from the early universe.  Their discovery would prove that standard field theory breaks down at short
distance scales.
\end{abstract}

\maketitle

\section{Introduction}

Ho\v{r}ava has suggested that nature may exhibit Lifshitz-like anisotropic scaling in the extreme ultraviolet (UV) regime
in order to cure the divergence problem in quantum gravity \cite{Horava, Melby}.
If gravity were to behave this way, then presumably all
other interactions should also.  In this paper, we shall show that Lifshitz scalar field theories can support stable,
finite-energy soliton or defect solutions of a kind that are impossible in standard second-order Lorentz-invariant
field theories.  Specifically, we shall demonstrate that stable, static, finite-energy defects formed entirely from
scalar fields can exist in three dimensions.  We shall derive both point-like and string-like solitons, which we can
call ``Lifshitz poles (L-poles)'' and ``Lifshitz strings (L-strings)'', respectively.
The non-existence of solutions having all of the above characteristics in standard field theories
follows from Derrick's theorem \cite{Derrick}.
So, while point-like (``hedgehog'') and global string solutions do exist in standard theory,
they have divergent energies or linear energy densities, as the case may be.
The discovery of L-poles and/or L-strings as cosmological relics from the early universe would
provide observational proof of the breakdown of standard field theory at short distances.

The basic idea of Lifshitz theories is to add high-order spatial derivative terms to the action while maintaining time
derivatives at second order.  This procedure preserves unitarity, while softening UV divergences through the presence
of higher powers of 3-momentum in the denominators of propagators.   This is an interesting idea, because it increases
the range of power-counting renormalisable field theories.  Ho\v{r}ava gravity \cite{Horava} is the best-studied
example; field-theory models involving extra dimensions of space is another \cite{Thompson}.
The improved UV behaviour comes
at the expense of Lorentz invariance, with the UV isotropic scaling of Lorentz-invariant field theories replaced by
anisotropic scaling between space and time.  This is obviously the most serious drawback of this approach, in both
a phenomenological and an aesthetic sense.  It is not known how to ensure that standard relativity emerges in the
infrared, except by fine-tuning, though progress has recently been made on the gravity front \cite{Melby}.
More generally, it is neither
clear how to ensure that all particle species acquire a common limiting speed, nor how to make the renormalisation
group running of the limiting speed(s) phenomenologically acceptable \cite{Iengo}.
So, there are arguments both for and against
Lifshitz theories of particle physics and gravity.  Experimental
or observational evidence in favour of the breakdown of Lorentz invariance in the UV would lend support to the
Lifshitz approach.  Such evidence can be searched for in very high energy cosmic ray events, but also through the
possible existence of cosmological relics that would be inexplicable by standard field theory.

We shall examine Lifshitz theories having fourth-order spatial derivatives, which is the simplest possible choice
from the point of view of the differential equations.  In general, these theories are much more difficult to analyse than
the usual second-order relativistic theories.  However, a Lifshitz analogue of the Bogomolnyi-Prasad-Sommerfield
(BPS) \cite{BPS}
superpotential approach can be invoked to reduce the problem back to second-order (this is equivalent to the "detailed balance" condition \cite{Horava:2008jf}).  We shall take this approach
in this paper, partly for simplicity, and partly because the BPS system has intrinsic theoretical interest through a deep
connection with supersymmetry and stochastic quantisation.

In the next section we review Lifshitz scalar field theory and anisotropic scaling.  We then show why Derrick's theorem
is inapplicable.  This is followed by the derivation of Lifshitz BPS soliton solutions, and then a brief section
on the general fourth-order system.  The final section is a conclusion.

\section{Lifshitz scalar field theory}

We shall work exclusively in $3+1$ dimensions. The construction of a Lifshitz theory begins through the specification
of the dynamical critical exponent $z$,
\begin{equation}
\vec{x} \to \kappa \vec{x}, \qquad t \to \kappa^z t,
\end{equation}
which defines the anisotropic scaling invariance of the theory in the UV.\footnote{We shall not consider the possibility
that different dimensions of space may scale differently.}   Lorentz-invariant theories must have isotropic
scaling, namely $z=1$.  Consider now the case of $z=2$, adopted for definiteness and simplicity in our analysis. We
shall construct the action step-by-step.  One first chooses that the Lagrangian contains only the term $\dot{\phi}^2$
in time derivatives.  This is to ensure the unitarity of the quantum theory.  Anisotropic $z=2$ scale invariance then follows
with the observation that the action $\int d^3x\, dt\, \dot{\phi}^2$ is invariant under
\begin{equation}
\vec{x} \to \kappa \vec{x},\quad t \to \kappa^2 t,\quad \phi(\vec{x}, t) \to \kappa^{-1/2} \phi(\vec{x}, t),
\label{eq:zeq2scaling}
\end{equation}
or, equivalently, under the field substitution
\begin{equation}
\phi(\vec{x}, t) \to \kappa^{+1/2} \phi(\kappa \vec{x}, \kappa^2 t),
\end{equation}
without the coordinates in the integration measure and derivatives participating in the transformation.  We shall adopt
the viewpoint of Eq.~(\ref{eq:zeq2scaling}) for definiteness in this paper.

The $\dot{\phi}^2$ action is actually invariant for any $z$, with the field transforming as
$\phi \to \kappa^{\frac{z-3}{2}} \phi$.  The $z$ value is specified through the spatial-derivative terms we now introduce.
The relativistic choice, $\int d^3x\, dt\, (\dot{\phi}^2 - c^2 \nabla \phi \cdot \nabla \phi)/2$, enforces $z=1$.  But we are
interested in $z=2$, which implies that
\begin{equation}
S \subset \int d^3x\, dt\,  \frac{1}{2} \left[ \dot{\phi}^2 - (\nabla^2 \phi)^2 \right].
\end{equation}
The arbitrary constant in front of the second term has been absorbed into $\vec{x}$, and the minus
sign is necessary to make the energy bounded from below.\footnote{Note that the constant
must be reintroduced if one wishes to work in standard natural units.}  The higher spatial derivative implies that the
propagator behaves like $1/[k_0^2 - (\vec{k} \cdot \vec{k})^2]$, so has improved UV behaviour compared to the
relativistic case $1/(k_0^2 - \vec{k} \cdot \vec{k})$.

The most general $z=2$ scale invariant action for a single real scalar field is then easily deduced to
be\footnote{Note that we are assuming that surface terms may be discarded,
so that there is no need to include $\phi^4 \nabla \phi \cdot \nabla \phi$, $\nabla \phi \cdot \nabla(\nabla^2 \phi)$ and
$\phi \nabla^2(\nabla^2 \phi)$ as independent terms.}
\begin{equation}
S = \int d^3x\, dt\,  \left[ \frac{1}{2} \dot{\phi}^2 - \frac{1}{2} (\nabla^2 \phi)^2  - a \phi^5 \nabla^2 \phi -
\lambda \phi^{10} \right],
\label{eq:scaleinvS}
\end{equation}
where $a$ and $\lambda$ are coupling constants, and they are dimensionless with respect to the ``anisotropic
scaling units'',
\begin{equation}
[S]=0,\qquad [x]=-1,\qquad [t]=-2,\qquad [\phi]=\frac{1}{2},
\end{equation}
that replace the usual ``mass dimension units'' of relativistic theories.  The power-counting renormalisation properties
follow from the adoption of these new units, with the usual rule that renormalisable terms have either dimensionless
or positive-power coupling constants, corresponding to marginal and relevant operators, respectively.  The scale-invariant
action (\ref{eq:scaleinvS}) contains only marginal operators.  The most general power-counting renormalisable action
contains in addition the terms,
\begin{eqnarray}
& \phi^n \nabla^2 \phi,\quad {\rm with}\quad 0 \le n \le 4 & \label{eq:phinablaphiterms} \\
& \phi^m,\quad {\rm with}\quad 0 \le m \le 9. \label{eq:Vterms}
\end{eqnarray}
These relevant operators explicitly, but softly, break the anisotropic scale invariance.
The most general action may be compactly written as
\begin{equation}
S = \int d^3x\, dt\,  \left[ \frac{1}{2} \dot{\phi}^2 - \frac{1}{2} (\nabla^2 \phi)^2  - K(\phi) \nabla^2 \phi - V(\phi) \right],
\label{eq:LifshitzS}
\end{equation}
where $K(\phi)$ is quintic and the potential $V(\phi)$ is tenth-order.  Observe that the standard relativistic term
$\nabla \phi \cdot \nabla \phi$ is contained as the $n=1$ term of $K$, up to a discarded total derivative.
The equation of motion that follows from Hamilton's principle is
\begin{equation}
\ddot{\phi} + \nabla^2(\nabla^2 \phi) - K_{\phi\phi} \nabla \phi \cdot \nabla \phi - 2 K_{\phi} \nabla^2\phi
+ V_{\phi} = 0,
\label{eq:Eqofmotion}
\end{equation}
where the subscripts $\phi$ and $\phi\phi$ denote first and second derivatives with respect to $\phi$, respectively.
The energy functional is
\begin{equation}
E[\phi] = \int d^3x \left[ \frac{1}{2} \dot{\phi}^2 + \frac{1}{2} (\nabla^2 \phi)^2 + K(\phi) \nabla^2 \phi + V(\phi) \right].
\label{eq:Egeneral}
\end{equation}
The generalisation of the above formalism to more than one scalar field is obvious.

\section{Derrick's theorem and the virial relation}

Derrick proved that stable, finite-energy, static solutions to second-order
relativistic scalar field theories do not exist in greater than two spatial dimensions \cite{Derrick}.
We now review his extremely simple argument for the case of three spatial dimensions.  For static configurations,
extremising the action functional is equivalent to extremising the energy functional,
\begin{equation}
E[\phi] = \int d^3x \left[ \frac{1}{2} \nabla \phi \cdot \nabla \phi + V(\phi) \right].
\end{equation}
Consider the specific field variation given by
\begin{equation}
\phi(\vec{x}) \to \phi(k \vec{x}),
\label{eq:Derrickvar}
\end{equation}
under which the energy varies to
\begin{equation}
E_k[\phi] = \int d^3x \left[ \frac{1}{2} \nabla \phi(k \vec{x}) \cdot \nabla \phi(k \vec{x})
+ V\left(\phi(k \vec{x})\right) \right].
\end{equation}
A necessary condition for a solution is that the variation of the energy under this restricted field variation should vanish
to first order, which is equivalent to the demand that
\begin{equation}
\left. \frac{\partial E_k[\phi]}{\partial k} \right|_{k=1} = 0.
\label{eq:firstvar}
\end{equation}
Now,
\begin{eqnarray}
E_k[\phi] & = & \int d^3\xi \left[  k^{-1} \frac{1}{2} \nabla_{\vec{\xi}}\, \phi(\, \vec{\xi}\, ) \cdot \nabla_{\vec{\xi}}\,
\phi(\, \vec{\xi}\, )  + k^{-3} V\left(\phi(\, \vec{\xi}\, )\right) \right], \nonumber\\
& \equiv & k^{-1} S_1 + k^{-3} S_3
\end{eqnarray}
where $\vec{\xi} \equiv k \vec{x}$ and $E = S_1 + S_3$.  Imposing Eq.~(\ref{eq:firstvar}) we get the virial relation
$-S_1 = 3 S_3$, which implies that $E = 2 S_1/3$.  A stable solution must have a positive second-order variation.
Computing $d^2 E_k/dk^2|_{k=1}$ one obtains $2S_1 + 12 S_3$, which equals $-2S_1$ through the virial theorem.
Since $S_1$ is positive-definite, no stable solutions can exist. It was implicit above that $E[\phi]$ converged, so the
theorem does not preclude infinite-energy, static, point-like solitons (the hedgehog \cite{Polyakov}
being a well-known example).

Repeating this exercise for the Lifshitz energy functional (\ref{eq:Egeneral}) we obtain
\begin{equation}
E_k[\phi] = k S_1 + k^{-1} S_2 + k^{-3} S_3,
\end{equation}
where $S_1 \equiv \int d^3x\, dt\, (\nabla^2 \phi)^2/2$ and
$S_2 \equiv \int d^3x\, dt\, K(\phi) \nabla^2 \phi$, which implies the virial relation
\begin{equation}
S_1 - S_2 - 3S_3 = 0.
\end{equation}
The second-order variation is now equal to $2 S_2 + 12 S_3$, which in turn equals $4 S_1 - 2 S_2$ through the virial
relation.  There is no necessity for this quantity to be negative, hence there is no generalisation of Derrick's theorem
to the $z=2$ Lifshitz case.  A simple example is the case $K=0$ which implies $S_2=0$ and hence a positive
second-order variation.  Another is $K = \phi^5$, which sees
$S_2$ turn into an integral of the negative-definite function $- \phi^4 \nabla \phi \cdot \nabla \phi$
after integration by parts.  This makes
$4 S_1 - 2 S_2$ necessarily positive.  The addition of higher-derivative terms thus tends to stabilise
scalar-field solitons, a phenomenon also observed in the (relativistic) Skyrme model \cite{Skyrme}.
However, evasion of Derrick's
theorem does not prove that stable, static, finite-energy, scalar-field solitons actually exist, because the variation
(\ref{eq:Derrickvar}) is not general.  We therefore turn to explicit computations.

\section{The Lifshitz BPS soliton solutions}

A technical problem with $z=2$ Lifshitz theory is that the scalar field equations are fourth-order in spatial derivatives,
as per Eq.~(\ref{eq:Eqofmotion}).  To get to concrete results in the simplest possible way, we therefore develop
a Lifshitz generalisation of superpotential systems.  This will reduce the problem to second order, and as a bonus
the solutions will be Lifshitz generalisations of BPS states, which is interesting in itself because of the
strong connection with supersymmetry and stochastic quantisation.  We shall derive three kinds of solutions:
a perturbatively-stable non-topological point-like soliton, a topological point-like soliton or hedgehog, and a topological
string.  In the section after this, we shall also briefly discuss the general fourth-order problem,
and display a finite-energy solution by way of an existence proof, but defer a systematic treatment to future work.

We first very briefly review how superpotential considerations lead to interesting simplifications in
one-dimensional second-order systems, with
\begin{equation}
E[\phi] =  \int dx \left[ \frac{1}{2} \left( \frac{d\phi}{dx} \right)^2 + V(\phi) \right].
\end{equation}
For the special cases where
\begin{equation}
V(\phi) = \frac{1}{2} \left( \frac{dW(\phi)}{d\phi} \right)^2,
\end{equation}
with $W$ being a superpotential, it is easy to see that
\begin{equation}
E[\phi] = \mp \left[ W\left(\phi(x = +\infty)\right) - W\left(\phi(x = -\infty)\right) \right]
+ \int_{-\infty}^{+\infty} dx\, \left( \frac{d\phi}{dx} \pm \frac{dW}{d\phi} \right)^2.
\end{equation}
Since the integrand above is positive definite, solutions to the first-order equations
\begin{equation}
\frac{d\phi}{dx} \pm \frac{dW}{d\phi} = 0
\end{equation}
globally minimise the energy functional for given boundary conditions.  They are stable BPS solutions,
and it is trivial to confirm that they satisfy the Euler-Lagrange equations.

We can achieve a situation that resembles the above if we restrict our Lifshitz action functional to the special
form
\begin{equation}
S = \int d^3x\, dt\, \left[ \frac{1}{2} \dot{\phi}^2 - \frac{1}{2} \left( \nabla^2\phi + K(\phi) \right)^2 \right],
\label{eq:superS}
\end{equation}
so that
\begin{equation}
V(\phi) = \frac{1}{2} K(\phi)^2.
\end{equation}
A superpotential $W$ can then be defined as the negative
indefinite integral of $K$ with respect to $\phi$, so that $K = - W_\phi$ (the minus sign is just a convention).
The convenience of the action (\ref{eq:superS}) is that the energy functional,
\begin{equation}
E[\phi] = \int d^3x \, \left[ \frac{1}{2} \dot{\phi}^2 + \frac{1}{2} \left( \nabla^2\phi + K(\phi) \right)^2 \right],
\end{equation}
is positive definite.  Configurations that obey
\begin{equation}
\dot{\phi} = 0,\qquad \nabla^2 \phi + K(\phi) = 0
\label{eq:LifshitzBPS}
\end{equation}
minimise the energy, and in fact they have zero energy-{\it density}.  It is straightforward to confirm that solutions
of Eq.~(\ref{eq:LifshitzBPS}) are also solutions of the field equation (\ref{eq:Eqofmotion}).  We shall call these
``Lifshitz BPS'' or LBPS solutions.

From Eq.~(\ref{eq:Eqofmotion}), we see that static, spatially homogeneous solutions (vacua $\phi = v$) must obey
$V_\phi(v) = 0$ as usual.  For a Lifshitz superpotential system, this means that
$K(v) K_\phi(v) = 0$, which implies that either $K(v) = 0$, or $K_\phi(v) = 0$ with $K(v) \neq 0$.
The former type of vacuum obeys
Eq.~(\ref{eq:LifshitzBPS}), while the latter does not.  Suppose $K(\phi) \neq 0\ \forall \phi$.  Then LBPS solutions
must obey either $\nabla^2 \phi > 0$ everywhere, or $\nabla^2 \phi < 0$ everywhere.  In either case, these solutions will
be asymptotically unbounded, and thus forbidden once physically reasonable boundary conditions are imposed.
We shall not consider such theories any further, by requiring that $K$ vanishes
for all vacuum configurations.  This means the LBPS solutions for the allowed models are degenerate with the vacua,
and hence at least perturbatively stable.

An interesting aspect of Lifshitz theories with $z=2$ anisotropic scaling (\ref{eq:superS}) is that they provide an
equivalent description of certain stochastic systems \cite{ParisiWu, ParisiSourlas}
(for recent relevant works see \cite{Dijkgraaf:2009gr} and references therein).
It is well known that the equilibrium limit --- defined
as the $\tau \to \infty$ limit where $\tau$ is a fictitious time (Markov parameter) ---
of the statistical correlators of
a $(d+1)$-dimensional scalar field $\phi_{\eta}(\tau,x^i)$ is equivalent to quantum correlators of
the $d$-dimensional\footnote{In what follows we explicitly discuss the d=3 case.} Euclidean scalar
field $\phi (x^i)$ with the classical action $S_{\rm E}[\phi]$:
\begin{equation}
\lim_{\tau\to \infty} \langle \phi_{\eta}(\tau, x_1^i)...\phi_{\eta}(\tau, x_n^i) \rangle_{\eta} =
\langle \phi(x_1^i)...\phi(x_n^i) \rangle.
\label{a1}
\end{equation}
Here $\phi_{\eta}(\tau,x^i)$ is a stochastic field, being a solution of the Langevin equation,
\begin{equation}
\partial_{\tau} \phi(\tau, x^i)=-\overline{\frac{\delta S_{\rm E}}{\delta \phi}}+\eta(\tau, x^i),
\label{a2}
\end{equation}
where
\begin{equation}
\overline{\frac{\delta S_{\rm E}}{\delta \phi}}\equiv \left. \frac{\delta S_{\rm E}}{\delta \phi(x^i)} \right |_{\phi(x^i) =
\phi(\tau, x^i)}
\end{equation}
 and $\eta(\tau, x^i)$ is Gaussian white noise, i.e.,
 \begin{equation}
 \langle \eta(\tau', x'^i)\eta(\tau, x^i)\rangle_{\eta}=\delta(\tau'-\tau)\delta^3(x'^i-x^i),
 \label{a3}
 \end{equation}
with all other correlators being zero. Note that in the large time limit, $\tau \to \infty$, the system approaches
equilibrium with $\partial_\tau \phi(\tau, x^i)=\eta=0$, and the Langevin equation (\ref{a2}), reduces to the
equation of motion of the Euclidean scalar field $\phi(x^i)$ described by the action $S_{\rm E}$.

The generating functional for the statistical correlators in Eq.~(\ref{a1}) reads as:
\begin{equation}
Z[J]=\int D\eta \exp \left \lbrace \int d\tau d^3x\left[-\frac{\eta^{2}}{2}+J\phi_{\eta}\right] \right\rbrace,
\label{a4}
\end{equation}
and
\begin{equation}
\langle \phi_{\eta}(\tau,x_1^i)...\phi_{\eta}(\tau,x_n^i) \rangle_{\eta} \equiv
\frac{1}{Z}\left. \frac{\delta^n Z}{\delta J(\tau,x_n^i)...\delta J(\tau,x_1^i)}\right \vert_{J=0}.
\end{equation}
Following the standard steps, we perform change of variables $\eta (\tau,x^i) \to \phi_{\eta}(\tau,x^i)$
using Eq.~(\ref{a3}) and  express the new functional measure through a path integral over
the Grassmannian fields $\psi$ and $\bar \psi$:
\begin{eqnarray}
D \eta \to D\phi \det\left( \frac{\delta \eta}{\delta \phi}\right)=D\phi  \det\left (\partial_{\tau} +\overline{\frac{\delta^2 S_{\rm E}}{\delta \phi^2}}\right) \nonumber \\
=D\phi \int D\bar{\psi}D\psi \exp\left \lbrace  \int d\tau d^3x\bar{\psi}\left (\partial_{\tau}+\overline{\frac{\delta^2 S_{\rm E}}{\delta \phi^2}}\right)\psi\right \rbrace.
\label{a5}
\end{eqnarray}
Plugging (\ref{a5}) into (\ref{a4}) and using the Langevin equation, we obtain the effective action
\begin{equation}
S_{\rm eff}=\int d\tau d^3x \left[\frac{1}{2}(\partial_{\tau}\phi)^2 +
\frac{1}{2}\left(\overline{\frac{\delta S_{\rm E}}{\delta \phi}}\right)^2-\bar{\psi} \left (\partial_{\tau}+
\overline{\frac{\delta^2 S_{\rm E}}{\delta \phi^2}}\right)\psi \right],
\label{a6}
\end{equation}
where we have dropped the full time derivative term $\partial_{\tau}\phi \overline{ \frac{\delta S_{\rm E}}{\delta \phi}}$. Taking
\begin{equation}
S_{\rm E}= \int d^3x \left[\frac{1}{2}(\partial_i\phi)^2 + W(\phi) \right]
\label{a7}
\end{equation}
in (\ref{a6}) and passing to the Lorentzian time $t=- i\tau$, we obtain a theory described by the action:
\begin{equation}
S_{\rm LSUSY}=\int dtd^3x \left[\frac{1}{2}\dot{\phi}^2-\frac{1}{2}\left(\nabla^2\phi -
\frac{dW}{d\phi}\right)^2+\bar{\psi} \left (-i\partial_{t}-\nabla^2+\frac{d^2W}{d\phi^2}\right)\psi \right].
\label{a8}
\end{equation}
The bosonic part of the above action is precisely the action of the $z=2$ Lifshitz
theory (\ref{eq:superS}) with $K(\phi)=-W_{\phi}$.

The action (\ref{a8}) is invariant under rigid (quantum mechanical) supersymmetry transformations.
To obtain transformations that are linear in the fields, we rewrite (\ref{a8}) in the equivalent form
\begin{eqnarray}
S_{\rm LSUSY}=\int dtd^3x \left[\frac{1}{2}\dot{\phi}^2+\frac{1}{2}D^2-D\left(\nabla^2\phi -
\frac{dW}{d\phi}\right)+\bar{\psi} \left (-i\partial_{t}-\nabla^2+\frac{d^2W}{d\phi^2}\right)\psi \right],
\label{a9}
\end{eqnarray}
where $D(t,x^i)$ is an auxiliary field satisfying the non-dynamical equation of motion:
\begin{equation}
D=\nabla^2\phi -\frac{dW}{d\phi}.
\label{a10}
\end{equation}
The infinitesimal supersymmetry transformations then read:
\begin{eqnarray}
\delta_{\epsilon}\phi = \bar \epsilon \psi - \bar \psi \epsilon,\quad \delta_{\epsilon} D=-i\partial_t(\bar \epsilon \psi + \bar \psi \epsilon),\nonumber \\
\delta_{\epsilon}\psi=\epsilon (D-i\dot \phi), \quad \delta_{\epsilon}\bar \psi=-\bar\epsilon (D+i\dot \phi).
\label{a11}
\end{eqnarray}
It is clear now that any solution of $z=2$ Lifshitz theory that satisfies Eq.~(\ref{eq:LifshitzBPS}) gives vanishing
D-term (\ref{a10}). Therefore, our LBPS solutions are supersymmetric and, hence, they are degenerate with
the zero-energy ground state.  Supersymmetry ensures the stability of the LBPS solitons.

From the perspective of stochastic dynamics (\ref{a2}), our supersymmetric LBPS solitons are relevant to the
description of physics at equilibrium ($\tau \to \infty$), while any time-dependent solution
describes physics in the non-equilibrium regime.
A discussion of these interesting points is beyond the scope of the present paper.

\subsection{Non-topological BPS L-pole}

Our first explicit solution will be a non-topological BPS L-pole.  Let $\phi = \phi_0(r)$ be a radially symmetric
solution of Eq.~(\ref{eq:LifshitzBPS}):
\begin{equation}
\left( \frac{2}{r} \frac{d}{dr} + \frac{d^2}{dr^2} \right) \phi_0(r) = - K\left( \phi_0(r) \right).
\label{eq:radialLBPSeq}
\end{equation}
Impose the boundary condition that $\phi_0$ asymptotes to a finite value $v$ as $r \to \infty$.  Regularity at the origin
requires $\phi_0 \sim r^2$ as $r \to 0$.  An obvious interpolation between these behaviours is provided by the
function
\begin{equation}
\phi_0(r) = \frac{v r^2}{R^2 + r^2},
\label{eq:1stexample}
\end{equation}
where $R$ is a parameter that sets the characteristic size of the defect.  It is easy to verify that this function solves
Eq.~(\ref{eq:radialLBPSeq}) for
\begin{equation}
K(\phi) = \frac{2}{R^2 v^2} \left( 4 \phi - 3v \right) \left( \phi - v \right)^2,
\label{eq:K1stexample}
\end{equation}
so that $v$ is a vacuum.
This provides a simple analytical example of an L-pole.  It is non-topological, because the asymptotic
limit is always the same vacuum.

To get the simple analytical form of Eq.~(\ref{eq:1stexample}) we of course need a very specific function $K$.  If we vary
this function continuously from Eq.~(\ref{eq:K1stexample}), then the solution will vary continuously away from our
analytical example, with numerical computations being in general necessary.

\begin{figure}
\centering
\includegraphics[width=0.75\textwidth]{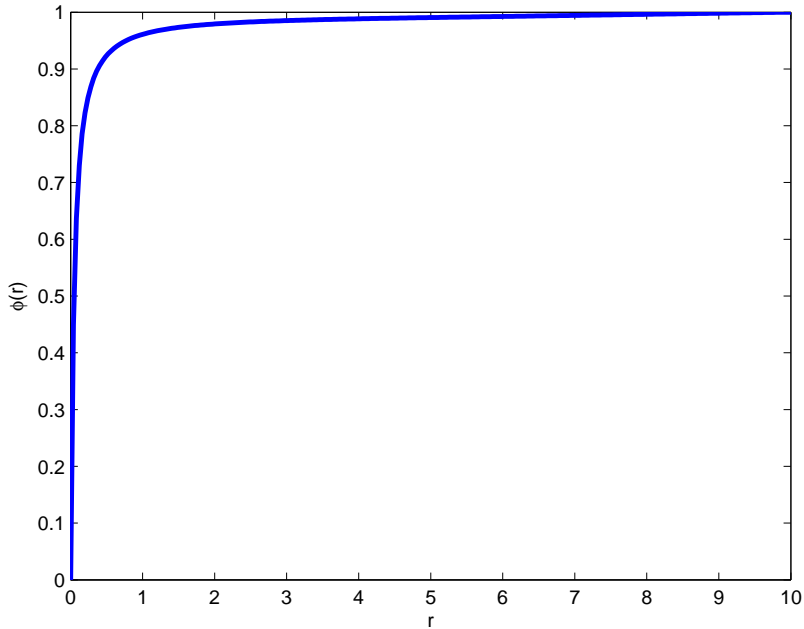}
\caption{Numerical solution for the non-topological BPS L-pole corresponding to the choice of $K$ given in Eq.~(\ref{eq:Knumerical}) with $\lambda = v = 1$.}
\label{Fig:non_top_BPS_L-pole}
\end{figure}

We give another example of a non-topological BPS L-pole which corresponds to a simpler choice of $K$, namely
\begin{equation}
K(\phi) = \lambda \phi (\phi^2 - v^2)^2.
\label{eq:Knumerical}
\end{equation}
The non-topological BPS L-pole corresponding to this choice of $K$ does not have an analytic solution,
however a numerical solution is shown in Fig.~\ref{Fig:non_top_BPS_L-pole}.

\subsection{Topological BPS L-pole or hedgehog}

\begin{figure}
\centering
\includegraphics[width=0.75\textwidth]{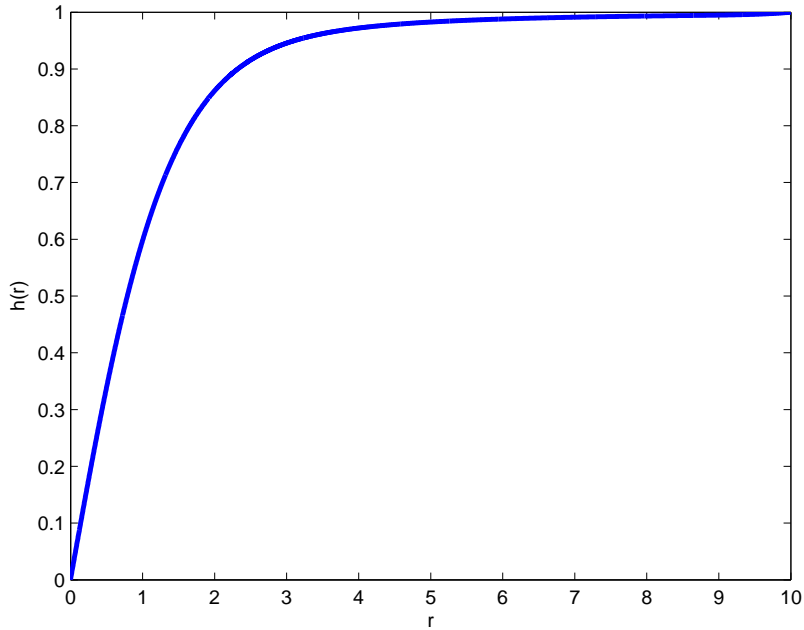}
\caption{Numerical solution to Eq.~(\ref{eq:h}) with $\lambda = v = 1$.}
\label{fig:LBPShedgehog}
\end{figure}

Our second example is an LBPS hedgehog.  The goal is to produce a topologically non-trivial BPS L-pole.  We need
the vacuum manifold to be a 2-sphere, so we choose an $O(3)$ invariant theory with a triplet of scalar
fields $\vec{\phi} = (\phi_i),\ i=1,2,3$, and work within the parameter regime that features
$O(3) \to O(2)$ spontaneous symmetry breaking.  The energy functional for static configurations is
\begin{equation}
E[\vec{\phi}] = \int d^3x\ \left[ \nabla^2 \phi_i + \phi_i F( \vec{\phi} \cdot \vec{\phi} ) \right]
\left[ \nabla^2 \phi_i + \phi_i F( \vec{\phi} \cdot \vec{\phi} ) \right],
\label{eq:EO3}
\end{equation}
where there is a sum over $i$ and
\begin{equation}
F = \lambda ( \vec{\phi} \cdot \vec{\phi} - v^2 )^2.
\end{equation}
The hedgehog ansatz in spherical polar coordinates is
\begin{equation}
\phi_i(r,\theta,\varphi) = \hat{x}_i h(r)
\label{eq:hedgehogansatz}
\end{equation}
where $(\hat{x}_1,\hat{x}_2,\hat{x}_3)$ is a unit vector in the $(\theta,\varphi)$ direction.  The LBPS equations
$\nabla^2 \phi_i + \phi_i F = 0$ reduce to
\begin{equation}
\frac{d^2h}{dr^2} + \frac{2}{r} \frac{dh}{dr} - 2 \frac{h}{r} + \lambda h (h^2 - v^2)^2 = 0,
\label{eq:h}
\end{equation}
which is the same as the original hedgehog equation \cite{Polyakov}
except that the non-derivative term is fifth-order rather
than third-order.  We demand that it asymptote to $v$ and be regular at the origin.
A numerical solution is depicted in Fig.~\ref{fig:LBPShedgehog}.

\subsection{Topological BPS L-string}

\begin{figure}[H]
\centering
\includegraphics[width=0.75\textwidth]{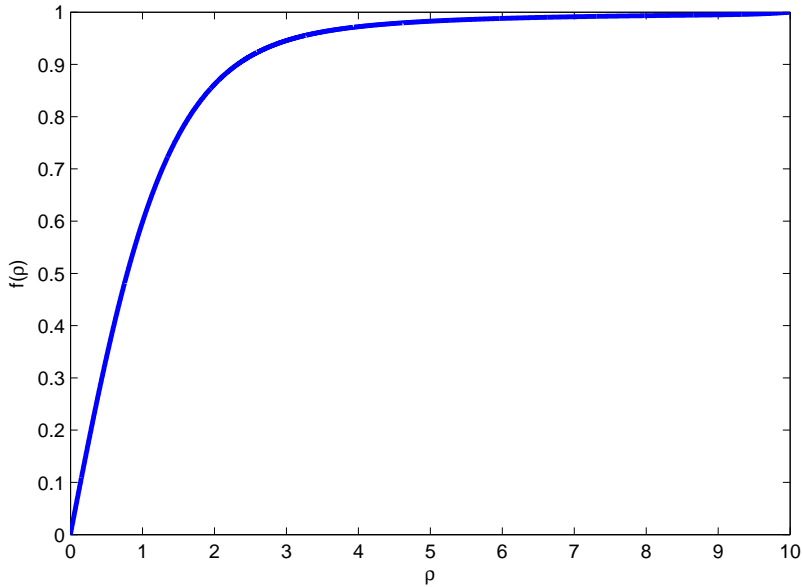}
\caption{Numerical solution to Eq.~(\ref{eq:f}) with $\lambda = v = 1$.}
\label{fig:LBPSstring}
\end{figure}

The third solution is a topological BPS L-string.  The theory has a complex scalar field $\Phi$ and a spontaneously
broken $\Phi \to e^{i\alpha} \Phi$ global $U(1)$ symmetry, with energy functional:
\begin{equation}
E[\vec{\Phi}] = \int d^3x\ \left[ \nabla^2 \Phi^* + \lambda \Phi^* ( \Phi^* \Phi - v^2 )^2 \right]
\left[ \nabla^2 \Phi + \lambda \Phi ( \Phi^* \Phi - v^2 )^2 \right].
\label{eq:Estring}
\end{equation}
The topological global string ansatz is
\begin{equation}
\Phi(\rho,\theta) = e^{i\theta} f(\rho)
\end{equation}
where $(\rho,\theta,z)$ are the usual cylindrical coordinates, and the string defines the $z$-axis. The BPS
equation of motion $\nabla^2 \Phi + \lambda \Phi ( \Phi^* \Phi - v^2 )^2 = 0$ reduces to
\begin{equation}
\frac{d^2 f}{d\rho^2} + \frac{1}{\rho} \frac{df}{d\rho} -\frac{f}{\rho^2} + \lambda f (f^2 - v^2)^2 = 0.
\label{eq:f}
\end{equation}
The asymptotic boundary condition is $f \to v$ as $\rho \to \infty$, and we demand regularity at the origin.
A numerical solution to Eq.~(\ref{eq:f}) is given in Fig.~\ref{fig:LBPSstring}.

\section{General Lifshitz systems}

One may wonder if the existence of the finite-energy, static solutions displayed above relies on the special
superpotential form of the action.  The answer is no, which we shall establish by displaying a
numerical solution for a non-LBPS hedgehog.  A full analysis of non-LBPS solutions is beyond the scope of this paper,
in part because the stability of non-topological soliton solutions requires the development of a fourth-order linear
stability formalism.

\begin{figure}[H]
\centering
\includegraphics[width=0.75\textwidth]{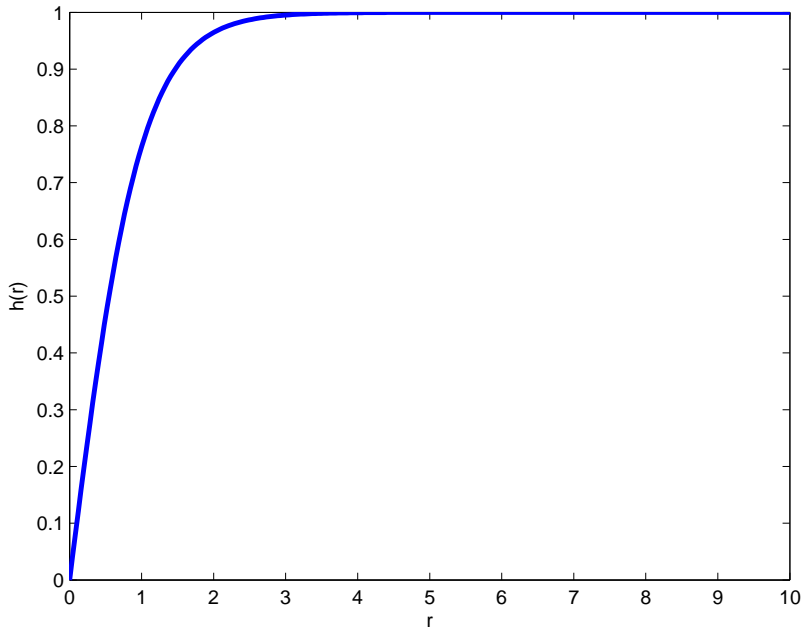}
\caption{Numerical solution to Eq.~(\ref{eq:nonBPShedgehogeqn}) with $\lambda = v = 1$.}
\label{fig:nonBPShedgehog}
\end{figure}

We shall exhibit a non-LBPS hedgehog solution for an $O(3)$ triplet model with $K=0$
and
\begin{equation}
V = \frac{\lambda}{10} \left( \vec{\phi} \cdot \vec{\phi} - v^2 \right)^5,
\end{equation}
where, using the ansatz of Eq.~(\ref{eq:hedgehogansatz}), the equation of motion reduces to
\begin{equation}
\frac{d^4 h}{dr^4} + \frac{4}{r} \frac{d^3 h}{dr^3} - \frac{4}{r^2} \frac{d^2 h}{dr^2} + \lambda h ( h^2 - v^2 )^4 = 0.
\label{eq:nonBPShedgehogeqn}
\end{equation}
A numerical solution for $h(r)$ is given in
Fig.~(\ref{fig:nonBPShedgehog}).  This configuration is topologically stable.

\section{Conclusions}

Lifshitz scalar field theories permit the existence of finite-energy, static soliton solutions that have only
infinite-energy analogues in standard second-order relativistic theories (such as the global string and the hedgehog).
We have displayed several such solutions, mostly within a superpotential-based subclass of $z=2$ Lifshitz
scalar theories that admit stable zero energy density BPS solutions.  These systems constitute the bosonic
sector of supersymmetric theories that are themselves related to a stochastic field theory in one higher dimension.
If the world is described by a Lifshitz theory, as suggested recently by Ho\v{r}ava as a way to improve ultraviolet
behaviour in a ghost-free manner, then the existence of new classes of finite-energy defects may provide a way
to look for observational evidence in support of this hypothesis.  To really make this connection, one would need to
extend our analysis from $z=2$ (which was chosen for simplicity), to at least $z=3$.  The gravitational properties
of Lifshitz solitons could then also be studied.

\section*{Acknowledgements}

We thank Damien George for useful comments.  This work was supported in part by the Australian Research Council
and in part by the Puzey bequest to the University of Melbourne.  RRV thanks O. Yasuda and H. Minakata for their
kind hospitality at Tokyo Metropolitan University where this work was completed.

\end{document}